\documentclass[%
 reprint,
 amsmath,amssymb,
 aps,
prx,
]{revtex4-2}

\usepackage{graphicx}
\usepackage{dcolumn}
\usepackage{amsthm}
\usepackage[caption=false]{subfig}
\usepackage{braket}
\usepackage{bm}
\usepackage{float}
\usepackage{xcolor}
\usepackage{url}
\usepackage{booktabs}
\usepackage[hidelinks]{hyperref}
\hypersetup{
  colorlinks   = true,
  urlcolor     = blue,
  linkcolor    = blue,
  citecolor   = blue
}
\usepackage[normalem]{ulem}

\newcommand\Yale{
    Departments of Applied Physics and Physics, 
    Yale University, 
    New Haven, CT 06520, USA
    }

\newcommand\YQI{
    Yale Quantum Institute, 
    Yale University, 
    New Haven, CT 06511, USA
    }

\begin{document}

\title{Untangling QLDPC Codes with Biased Noise Ancilla}

\author{Runjiang Bi}
\affiliation{\Yale}
\affiliation{\YQI}
\author{Kathleen (Katie) Chang}
\affiliation{\Yale}
\affiliation{\YQI}
\author{Shruti Puri}
\affiliation{\Yale}
\affiliation{\YQI}

\begin{abstract}
Remarkable technical progress has made high-rate, high-distance, quantum low-density parity-check codes (QLDPC) promising candidates for scalable quantum computing. However, it is hard to design low-depth syndrome extraction circuits that do not spread errors from ancilla qubits to multiple data qubits, also known as {\it hook errors}, for general QLDPC codes. Additionally, widely used decoders for these codes based on belief propagation are impaired due to short loops in the Tanner graph. Here, we investigate a hardware-aware approach to avoid these hooks and loops using biased noise ancillas. Using examples of bicycle bivariate codes and a cyclic hypergraph product code, which have been widely considered for practical application, we show that the effective fault-distance of the conventional syndrome extraction circuit can be significantly higher and the number of short loops can be significantly lower when the ancillas are subject to phase-flip errors only, compared to when they are also subject to bit-flip errors. This can result in almost an order of magnitude improvement in the logical error rate at circuit noise of $2\times 10^{-3}$ and when bit-flip errors in the ancilla are 50 times less likely than phase-flip errors. Our work demonstrates a significant and practical quantum error correction advantage with biased noise qubits in which full-bias cannot be maintained.
\end{abstract}
\date{\today}

\maketitle

\section{Introduction}

Fault-tolerant quantum error correction (QEC) is key for overcoming noise in quantum hardware to achieve scalable quantum computation. It is desirable to have {\it good} quantum error correction codes which have both high-rate and high-distance per physical qubit~\cite{panteleev2022asymptotically,leverrier2022quantum, dinur2023good}. Such good codes reduce the qubit overhead, however, they come at the cost of more complicated qubit connectivity. For example, the most widely-pursued surface code~\cite{kitaev2003fault,bravyi1998quantum, dennis2002topological} has a vanishing rate and distance in the limit of very large number of physical qubits, but its implementation requires only local qubit connectivity. On the other hand, several families of quantum low-density parity-check codes (QLDPC)~\cite{breuckmann2021quantum} have been constructed that outperform the surface code in terms of rate, but these require qubits with nonlocal connectivity. Nonetheless, with recent experimental advances towards realizing high-quality non-local couplings~\cite{bluvstein2024logical,evered2026high,moses2023race,heya2025randomized,qldpc2025cross,wang2026demonstration}, these codes are now becoming promising candidates for practical scalable quantum computation. For example, bicycle bivariate (BB) codes~\cite{bravyi2024high} and hypergraph product (HGP) codes~\cite{tillich2013quantum} have been studied for superconducting circuits~\cite{yoder2025tour}, neutral atoms~\cite{xu2025fast}, and trapped ions~\cite{aydin2025cyclic,tripier2026fault}. 

In practice, the performance of a QEC code is dependent on errors during stabilizer measurements and decoding accuracy. Typically, each stabilizer is measured using an ancilla and the circuits are designed to minimize the overall circuit depth. However, errors from the ancilla can propagate to multiple data qubits. This residual error is called a {\it hook error}. When the residual error has support on a minimum-weight logical operator, the effective distance of the code can be affected, and thus its logical performance degrades. Distance-preserving stabilizer measurement circuits are known for the surface code~\cite{tomita2014low} and HGP codes~\cite{manes2025distance}, but not for BB codes~\cite{bravyi2024high}, for example. 

Moreover, most QLDPC codes are decoded using variations of the belief propagation (BP) algorithm~\cite{panteleev2021degenerate, roffe2020decoding} on the Tanner graph representing the stabilizer checks of the code. However, due to short loops or cycles in the Tanner graph, these decoding strategies often fail, even after low-weight errors, thus undermining the logical performance of the code~\cite{poulin2008iterative,raveendran2021trapping}. Recent research has shown that the accuracy of decoding can be improved significantly if the underlying noise in the qubits is biased so that bit-flip errors are suppressed compared to phase-flips and if the QLDPC code is also appropriately tailored to leverage this noise bias~\cite{roffe2023bias, das2026clifford}.

However, there are some caveats to these previous results. Firstly, the conclusions are based on the availability of controlled-not (CX) gates that preserve the underlying noise asymmetry. Such gates cannot exist in conventional qubit platforms~\cite{aliferis2008fault,puri2020bias,Guillaud_2019} and are hard to implement even in those specially engineered platforms that can support such gates~\cite{puri2020bias, Guillaud_2019}. Secondly, the framework for tailoring the code is restricted to certain families of QLDPC codes. 

Remarkably, in this work, we demonstrate that it is possible to improve the performance of general QLDPC codes without any code tailoring, even when bias is not symmetrically distributed over all qubits and/or gates.  In particular, we find that restricting the bias in the ancilla qubit used in the stabilizer measurement circuit is sufficient to avoid hook errors and reduce the number of malignant short cycles, thereby improving the logical performance. In this work, the model where biased noise is restricted to ancilla qubits used for stabilizer measurements will be referred to as ancilla-only biased noise. Moreover, the model where all qubits are biased but a bias-preserving CX is unavailable will be referred to as non-CX bias. 

We support our arguments with numerical simulations of instances from the BB and HGP code families at experimentally realistic physical error rates.  We find the logical error rate of the [[144,12,12]] BB or gross code decreases 8.5-fold at circuit noise of $2\times 10^{-3}$ and when bit-flip errors in the ancilla are 50 times less likely than phase-flip noise. For the [[336,20,6]] cyclic HGP code, we find a 4.2-fold reduction in the logical error rate with the same noise parameters. In case of the HGP code, hook error-free stabilizer measurement circuits are known~\cite{manes2025distance}, so the improvement is due only to a decrease in the number of short cycles. Remarkably, we find that the non-CX bias has a worse logical performance than that under ancilla-only bias due to a higher rate of $Z$-type noise, but it still performs better than depolarizing noise.

The paper is structured as follows.
Section~\ref{sec:qldpc-codes-and-syndrome-extraction-circuits} describes the two families of QLDPC codes we study and the syndrome extraction circuits we use for our circuit-level simulations.
Section~\ref{sec:noise-models} describes and motivates the four different noise models we study, summarized in Table~\ref{tab:noise-model}.
Section~\ref{sec:hook} introduces what hook errors are and how biased ancilla noise can eliminate these errors. 
Section~\ref{sec:cycles} describes cycles on a Tanner graph, why they are harmful for decoding with BP, and how biased ancilla can mitigate cycles.
The results of our numerical studies are presented in Section~\ref{sec:results}.
Finally, we conclude in Section~\ref{sec:conclusion}.

\section{QLDPC codes and Syndrome Extraction Circuits}
\label{sec:qldpc-codes-and-syndrome-extraction-circuits}

\begin{figure*}
    \centering
    \subfloat[]{\includegraphics[width=0.5\linewidth]{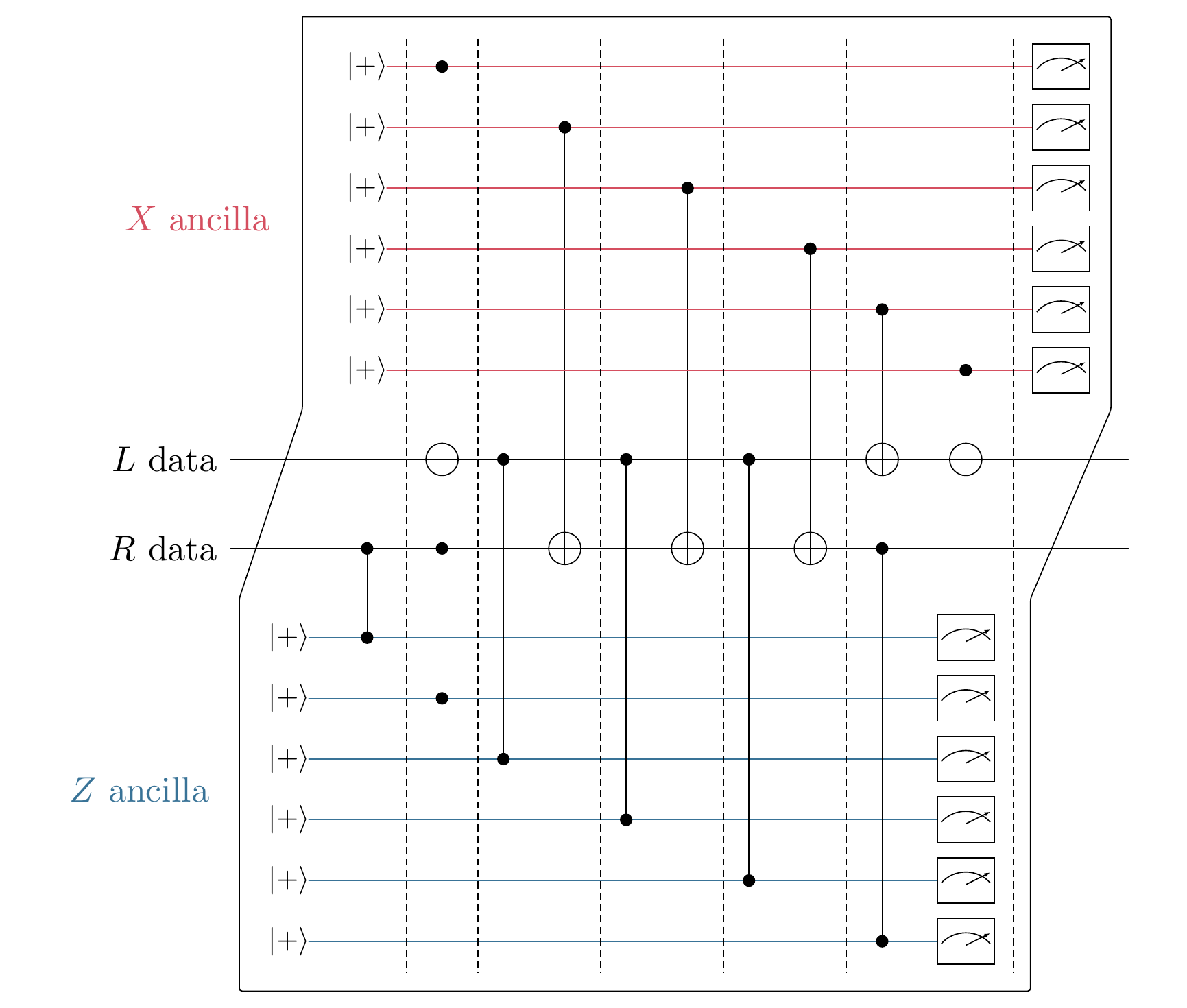}
    \label{fig:finite-bb}}
    \subfloat[]{
    \includegraphics[width=0.5\linewidth]{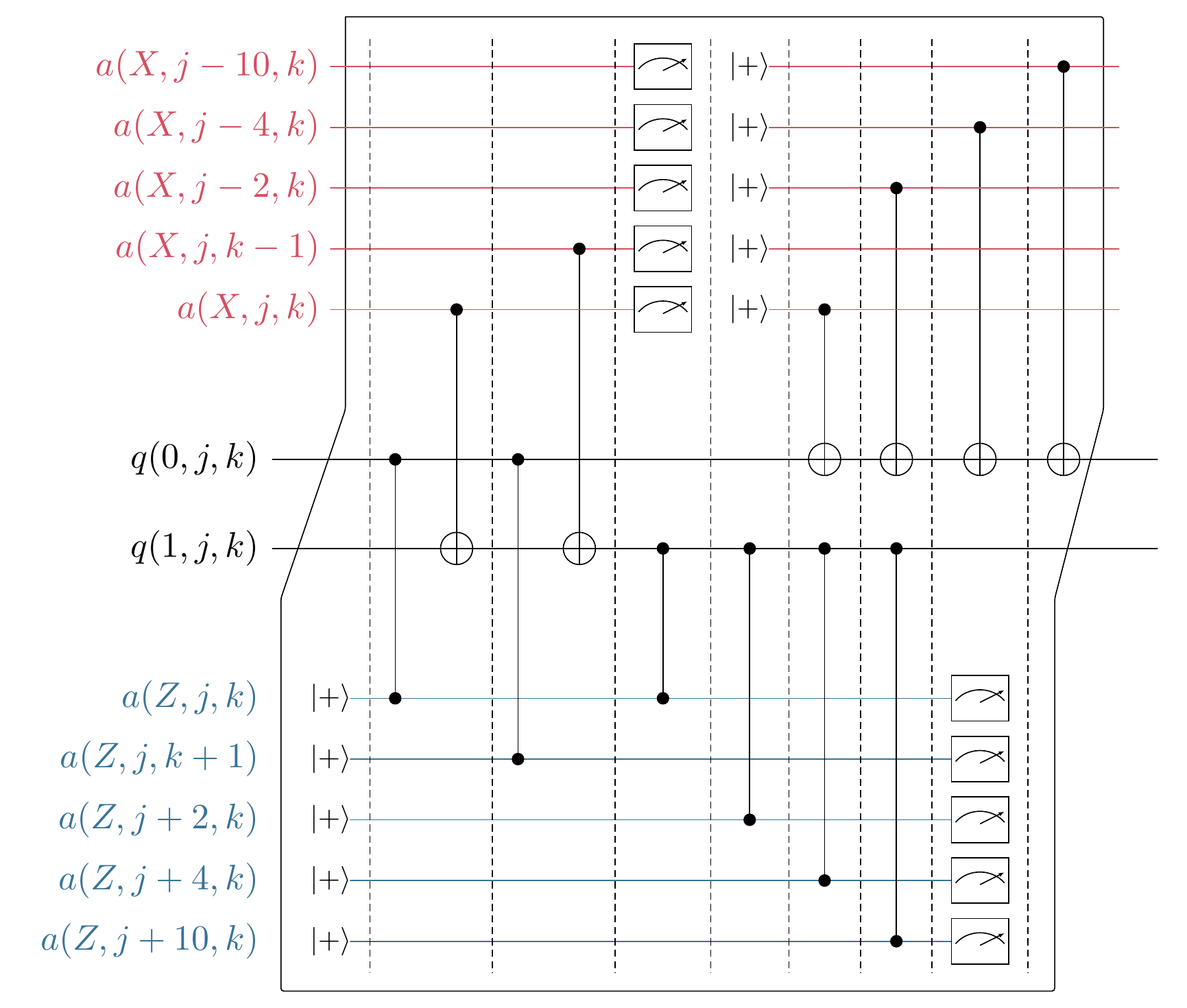}
    \label{fig:finite-hgp}}
    \caption{Syndrome extraction circuits. Data qubits are in black, $X$ ancilla qubits are in red, and $Z$ ancilla qubits are in blue. All ancilla qubits are measured in the $X$ basis. 
    \textbf{(a)} Syndrome extraction circuit for BB codes, from~\cite{bravyi2024high}. Data qubits are partitioned into left (L) data qubits and right (R) data qubits. Each data qubit interacts with three $X$-ancilla qubits and three $Z$-ancilla qubits in a depth-7 syndrome extraction cycle, where time-steps are partitioned by dashed lines.
   \textbf{(b)} Syndrome extraction circuit for the $[[336, 20, 6]]$ cyclic HGP code. Data qubits are indexed by $q(i,j,k)$, with $(i,j,k)\in \mathbb{Z}_2 \times \mathbb{Z}_{28} \times \mathbb{Z}_6$. Ancilla qubits are indexed by $a(r,s,t)$, with $(r,s,t)\in \{X,Z\} \times \mathbb{Z}_{28} \times \mathbb{Z}_6$. The circuit shows the connectivity for fixed data qubits $q(0,j,k)$ and $q(1,j,k)$ during one syndrome extraction round. Each ancilla label, $a(X,s,t)$ or $a(Z,s,t)$, indicates the $X$ or $Z$ ancilla qubit that couples to the corresponding data qubit. A modification of this circuit is first defined in~\cite{aydin2025cyclic}.}
    \label{fig:syndrome-extraction}
\end{figure*}

We study two families of QLDPC codes, BB codes and HGP codes.
Specifically, we benchmark three instances of BB codes with increasing distance, $[[72,12,6]]$, $[[90,8,10]]$ and the gross code, $[[144,12,12]]$, from Ref.~\cite{bravyi2024high}. The specific HGP code we study is the $[[336, 20,6]]$ cyclic hypergraph product code~\cite{aydin2025cyclic}. We study these codes specifically, as they are promising for practical implementation in many different hardware platforms~\cite{yoder2025tour, tham2025distributed, Poole_2025, Viszlai_2026}. The syndrome extraction circuits used are shown in Fig.~\ref{fig:syndrome-extraction}. Note that for all of the syndrome extraction circuits we study, each ancilla is always control of a CZ or CX gate, and it is always prepared and measured in the $X$-basis. This choice of two-qubit gates is motivated by the noise channels we study.

\section{Biased Noise Models}
\label{sec:noise-models}
\begin{table*}
\centering
\begin{tabular*}{\linewidth}{@{\extracolsep{\fill}} c c c c c}
\toprule
\textbf{Noise model} & Depolarizing & Ancilla-only bias &
Non-CX bias & Full-bias ($\eta=\infty$)\\
\midrule
SPAM &
$\mathsf{Zerr}_1[p]$ &
$\mathsf{Zerr}_1[p]$ &
$\mathsf{Zerr}_1[p]$ &
$\mathsf{Zerr}_1[p]$ \\

Idling &
$\mathsf{DEP}_1[p]$ &
$\mathsf{DEP}_1[p]$ &
$\mathsf{BIAS}_1[\eta,p]$ &
$\mathsf{Zerr}_1[p]$ \\

Two-qubit gates &
$\mathsf{DEP}_2[p]$ &
Eq.~\eqref{eq:anc-only-2q} &
CX: Eq.~\eqref{eq:non-bias-cx} &
$\mathsf{Zerr}_2[p]$ \\

 & & & 
CZ: $\mathsf{BIAS}_2[\eta,p]$ &  \\
\bottomrule
\end{tabular*}
\caption{
We present all four noise-models studied in this work: depolarizing, ancilla-only biased, non-CX biased, and full-biased noise models.
We delineate the error channels for state preparation and measurement (SPAM), idling, and two-qubit gate errors, all parameterized by physical noise strength $p$, and as relevant, the noise bias parameter $\eta$.
Note that SPAM error channels are inserted immediately before or after $X$ basis initialization and measurement of the ancilla qubits.
Two-qubit gate errors are inserted immediately following the gate. 
$\mathsf{DEP}_1[p]$ and $\mathsf{DEP}_2[p]$ are the single-qubit and two-qubit depolarizing error channels defined in Eqs.~\eqref{eq:1q-dep} and~\eqref{eq:2q-dep}. $\mathsf{Zerr_1}[p]$ and $\mathsf{Zerr_2}[p]$ single-qubit and two-qubit error channels which consist only of $Z$ errors, defined in Eqs.~\eqref{eq:1q-zerr} and~\eqref{eq:2q-zerr}. 
$\mathsf{BIAS_1}[\eta,p]$ and $\mathsf{BIAS}_2[\eta,p]$, defined in Eqs.~\eqref{eq:bias-1q} and~\eqref{eq:bias-2q}, are biased single-qubit and two-qubit error channels parameterized by $\eta$.} 
\label{tab:noise-model}
\end{table*}

In this section, we describe the four different noise models we study here which are also summarized in Table~\ref{tab:noise-model}. 
First, we consider an unbiased, depolarizing noise model. Under this model, the standard single-qubit depolarizing error channel, $\mathsf{DEP}_1[p]$, is applied to every idling qubit,  
\begin{align}
    \mathsf{DEP}_1[p](\rho) = (1-p)\rho + \frac{p}{3}(X\rho X + Y\rho Y + Z\rho Z)
    \label{eq:1q-dep}
\end{align}
Here, $p\in [0,1]$ is the physical noise strength parameterizing the error channel. A two-qubit depolarizing channel, $\mathsf{DEP}_2[p]$, is applied after every two-qubit gate to the two qubits involved in it,

\begin{align}
    \mathsf{DEP}_2[p](\rho) =(1-p)\rho + \frac{p}{15}\sum_{P_i\in S} P_i\ \rho\ P_i
    \label{eq:2q-dep}
\end{align}
with $S=\{I,X,Y,Z\}^{\otimes 2} \backslash\{II
\}$. We model state preparation and measurement error (SPAM) by applying a single-qubit $Z$ error channel,
\begin{align}\label{eq:1q-zerr}
    \mathsf{Zerr}_1[p](\rho) = (1-p)I\rho I + p Z\rho Z
\end{align}
\noindent right after (before) a noiseless \(\ket{+}\) state preparation (noiseless $X$ measurement) of the ancilla. 

Next, we describe the \textit{ancilla-only biased} noise model.
This model is motivated by heterogeneous hardware in which data and ancilla qubits may be encoded in different physical systems, with ancilla qubits dominated by dephasing noise and data qubits subject to depolarizing noise. The idling noise on the data qubit is $\mathsf{DEP}_1[p]$ as described in Eq.~\eqref{eq:1q-dep}.
We need not define ancilla idling noise, as there is no idling on the ancilla qubit for the syndrome extraction circuits in Fig.~\ref{fig:syndrome-extraction}.

Recall that the ancilla is always the control qubit in the two-qubit gates used for syndrome extraction. Such ancilla-controlled gates, in principle, can be designed using interactions that always commute with the underlying noise in the ancilla, thus preserving the ancilla noise bias. Therefore, the physically-motivated error channel we apply after every two-qubit gate is
\begin{align}
    \mathcal{N}[\eta, p](\rho) &= (1-p)\rho \notag\\
    &+ \frac{p\eta}{7\eta + 8}\biggl( \sum_{P_i \in S_1} IP_i\ \rho\ IP_i \ \notag\\
    &+\ \sum_{P_i\in S_2}
    Z P_i \rho Z P_i\biggr)\notag\\
    &+\frac{p}{7\eta + 8} \sum_{P_i\in S_3}\sum_{P_j\in S_2}P_iP_i\rho P_iP_j.
    \label{eq:anc-only-2q}
\end{align}
where the first Pauli operator in each term of the sum is applied to the ancilla qubit and the second is applied to the data qubit. Here, $S_1=\{X,Y,Z\}$, $S_2=\{I,X,Y,Z\}$, and $S_3=\{X,Y\}$.
The parameter $\eta$ represents the noise bias in the ancilla, and the choice of the channel ensures that when $\eta=1$, the channel in~\eqref{eq:anc-only-2q} reduces to $\mathsf{DEP}_2[p]$.
When $\eta\rightarrow\infty$, the noise channel has complete bias so that the probability of an $X$ or $Y$ error on the ancilla qubit is zero.

Another noise model we study is the \textit{non-CX biased} noise model, which is motivated by physical qubits where the underlying noise is biased towards dephasing errors but a bias-preserving CX gate is unavailable. Most qubits with biased noise, such as fluxonium qubits~\cite{pop2014coherent}, quantum dot spin qubits~\cite{shulman2012demonstration,watson2018programmable}, and nuclear spins in diamond~\cite{waldherr2014quantum}, fall under this category. In these systems, it is natural to be able to implement operations that commute with the dominant $Z$ error, such as a CZ gate, in a way that does not destroy the underlying bias. On the other hand, operations like CX, which do not commute with the underlying $Z$ noise, inherently destroy the bias. The noise in the CX gate can be described by the channel

\begin{align}
    \mathcal{N}[\eta, p](\rho) &= (1-p)\,\rho \notag \\
    &+ \frac{p\eta}{2\eta + 10}\Bigl(ZI\,\rho\,ZI \Bigr)\notag \\
    &+\frac{3p\eta}{16\eta + 80}\Bigl(IZ\,\rho\,IZ+ZZ\,\rho\,ZZ\Bigr) \notag\\
    &+\frac{p\eta}{16\eta + 80}\Bigl(IY\,\rho\,IY+ZY\,\rho\,ZY\Bigr)\notag \\
    &+ \frac{p}{2\eta + 10}
    \sum_{\substack{P_iP_j\in S_4 }}
    P_iP_j\,\rho\,P_iP_j .
\label{eq:non-bias-cx}
\end{align}
with $S_4=\{I,X,Y,Z\}^{\otimes 2}\backslash\{I,Z\}\otimes\{I,Z,Y\}$~\cite{darmawan2021practical}. As before, the first Pauli operator in each term of the sum is applied to the ancilla qubit and the second to data qubit. Using the same convention, the noise in the CZ gate is described by
\begin{align}
\mathsf{BIAS}_2[\eta,p](\rho)&= (1-p)\,\rho \notag \\
    &+ \frac{p\eta}{2\eta + 13}\Bigl(IZ\,\rho\,IZ+ZI\,\rho\,ZI \Bigr)\notag \\
    &+ \frac{p}{2\eta + 13}
   \sum_{\substack{P_iP_i\in S_5}}
   P_iP_j\,\rho\,P_iP_j .
    \label{eq:bias-2q}
\end{align}
with $S_5=\{I,X,Y,Z\}^{\otimes 2}\backslash\{II,IZ,ZI\}$.
The idling on data qubits is a single-qubit biased noise channel $\mathsf{BIAS}_1[\eta,p]$
\begin{align}
    \mathsf{BIAS}_1[\eta,p] (\rho) &= (1-p) \rho + \frac{p\eta }{(\eta +2)}Z\rho Z\notag \\
    &+  \frac{p}{(\eta + 2)}\left(X\rho X + Y\rho Y\right).
    \label{eq:bias-1q}
\end{align}
where $\eta$ represents the degree of suppression of $X,Y$ errors compared to $Z$ errors. When $\eta = 1$, this channel reduces to the single-qubit depolarizing channel $\mathsf{DEP}_1[p]$ and when $\eta\rightarrow\infty$, this channel approaches the pure-dephasing channel. As with the depolarizing model, SPAM is modeled by inserting  $\mathsf{Zerr}_1[p]$ right after and before $\ket{+}$ initialization and $X$-basis measurement, respectively.

Finally, we define a \textit{full-biased} noise model in which only $Z$ errors may occur.
This full-biased noise model is not meant to be a physically realizable target noise model, but it represents the limit of noise tailoring in which the bias is perfectly preserved at every circuit location.
In particular, all SPAM and idling errors are modeled by the $\mathsf{Zerr}_1[p]$ error channel.
After any two-qubit gate, the data and ancilla qubit undergo the following two-qubit $Z$ error channel,
\begin{align}
    \mathsf{Zerr_2}[p] &= (1-p)\rho \notag \\
    &+ \frac{p}{3}\left(IZ\rho IZ+ZI\rho ZI+ZZ\rho ZZ\right).
    \label{eq:2q-zerr}
\end{align}

\section{Hook Errors}
\label{sec:hook}

A hook error is a fault on a single ancilla qubit that, through the entangling gates applied after it, spreads to two or more data qubits.
If the residual error lies in the support of a minimum-weight logical operator of the code, then the \textit{circuit-level distance} of the code is reduced, and the logical performance degrades.

\begin{table}[t]
\centering
\caption{Upper bound of the circuit-level distance for three BB codes for all noise models. The ``Biased" column applies to all three biased noise models under infinite bias ($\eta\rightarrow \infty$): ancilla-only bias, non-CX bias, and full-bias. }
\label{tab:circuit_level_distance}
\begin{tabular*}
{\linewidth}{@{\extracolsep{\fill}} c c c c}
\toprule
Code & Depolarizing & Biased ($\eta \rightarrow\infty$) \\
 &  &  \textit{(all three models)} \\
\midrule
$[[72,12,6]]$ & 6 & 6 \\
$[[90,8,10]]$ & 8 & 10 \\
$[[144,12,12]]$ & 10 & 12 \\
\bottomrule
\end{tabular*}
\end{table}

For the syndrome extraction circuits we consider, each check is measured by a single ancilla qubit which is the control qubit for a set of CX or CZ gates. 
This convention fixes which Pauli errors on ancilla qubits can lead to hook errors.
Clearly, a $Z$ error on the ancilla commutes with all entangling gates, so this ancilla error never propagates onto multiple data qubits.
Conversely, an $X$ or $Y$ error on the ancilla qubit propagates onto the data qubits through the commutation relations with subsequent CXs or CZs.
It immediately follows that restricting the noise on ancilla qubits to only $Z$ errors eliminates hook errors from occurring. 

\begin{table*}[htbp!]
\centering
\caption{Number of $4$-cycles and $6$-cycles for the $[[144,12,12]]$ BB code under different noise models. $Z$-DEM and $X$-DEM are Tanner graphs with only $Z$ or $X$ detectors, respectively. The complete DEM contains both kinds of detectors.}
\label{tab:cycle-counts-bb}
\begin{tabular*}{\linewidth}{@{\extracolsep{\fill}} c c c c c}
\toprule
& Depolarizing & Ancilla-only bias & Non-CX bias & Full-bias \\
\midrule
$4$-cycles, $Z$-DEM & 53,280 & 5,184 & 0 & 0 \\
$4$-cycles, $X$-DEM & 52,416 & 5,184 & 5,184 & 5,184 \\
$4$-cycles, complete DEM   & 12,505,536 & 1,174,824 &185,688 & 79,704 \\
6-cycles, $Z$-DEM & 992,448 & 45,648 & 1,728& 0 \\
6-cycles, $X$-DEM & 971,424 & 45,648 & 45,648 & 45,648 \\
6-cycles, complete DEM   & 3,546,643,248 & 77,241,600 & 4,555,728 & 906,696 \\
\bottomrule
\end{tabular*}
\end{table*}

\begin{table*}[htbp!]
\centering
\caption{Number of $4$-cycles and $6$-cycles for the $[[336,20,6]]$ HGP code under different noise models. $Z$-DEM and $X$-DEM are Tanner graphs with only $Z$ or $X$ detectors, respectively. The complete DEM contains both kinds of detectors.}
\label{tab:cycle-counts-hgp}
\begin{tabular*}{\linewidth}{@{\extracolsep{\fill}} c c c c c}
\toprule
 & Depolarizing & Ancilla-only bias & Non-CX bias & Full-bias  \\
\midrule
4-cycles, $Z$-DEM & 222,768 & 25,872 & 1,008 & 0 \\
4-cycles, $X$-DEM & 188,496 & 25,872 & 25,872 & 25,872\\
4-cycles, complete DEM  & 29,087,856 & 1,511,664  & 313,152 & 121,968\\
6-cycles, $Z$-DEM & 8,870,400 & 457,800  & 12,096 & 0  \\
6-cycles, $X$-DEM & 7,656,432 & 457,800  & 457,800 & 457,800 \\
6-cycles, complete DEM   & 14,000,854,560 & 119,077,056 & 11,445,504 & 2,513,448 \\
\bottomrule
\end{tabular*}
\end{table*}

We confirm that there is indeed no reduction in the circuit-level distance for the biased noise models by numerically estimating an upper bound for the circuit distance for the $[[72, 12, 6]]$, $[[90,8,10]]$, and $[[144,12,12]]$ BB codes. The upper bound is found using BP to randomly search for the minimum-weight configuration of faults that anticommute with a logical and leave every detector unflipped, following the method described in Section 6 of~\cite{bravyi2024high}. The results are summarized in Table~\ref{tab:circuit_level_distance}, which also lists the upper bound found for the depolarizing noise model for reference.
Under depolarizing noise, the upper bound of the circuit-level distance for $[[90,8,10]]$ and $[[144,12,12]]$ is $\le 8$ and $\le 10$ respectively, indicating the existence of a distance-reducing hook error.
For all three biased noise models--- full-bias, non-CX bias, and ancilla-only bias--- the circuit-level distance upper bound is restored to code distances $10$ and $12$ for the gross and $[[90,8,10]]$ codes, respectively. 
This suggests that biasing the ancilla is sufficient to preserve the code distance in the syndrome extraction circuit and, therefore, could improve the logical performance compared to that under depolarizing noise.
The $[[336,20,6]]$ cyclic HGP code already has distance-preserving syndrome extraction circuit~\cite{manes2025distance}.
Consequently, we expect that improvements in its performance under biased noise arise due to reduction in short cycles, discussed in the next section.

\section{Cycles}
\label{sec:cycles}

Widely used decoding algorithms for QLDPC codes are based on BP which relies on iterative message passing on the Tanner graph of the circuit's detector error model (DEM)~\cite{gidney2021stim}. 
The check nodes in this graph are \textit{detectors} which are products of measurement outcomes that equal $+1$ in the absence of faults. For example, the product of measurements of the same stabilizer in two adjacent syndrome extraction rounds.
Each bit node represents an error mechanism, or a set of Pauli faults in the circuit, that flip the same set of detectors and logicals.

Generally, BP does not work well on Tanner graphs that contain short loops or cycles, a feature that quantum codes necessarily have~\cite{poulin2008iterative,raveendran2021trapping} and that gets amplified when circuit-level noise is considered. By reducing the number of independent error mechanisms, we expect to reduce the number of cycles. Thus, we expect the biased noise models to have a less loopy Tanner graph than the depolarizing noise model and therefore admit more accurate BP-based decoding.

We confirm the reduction in the number of short cycles in the biased noise models at $\eta\rightarrow \infty$ compared to depolarizing noise by exhaustively counting the number of $4$-cycles and $6$-cycles in the Tanner graph for the gross code and $[[336,20,6]]$ HGP code. The results are presented in Table~\ref{tab:cycle-counts-bb} and \ref{tab:cycle-counts-hgp}. We compute these counts for three DEMs, one with only $X$- detectors, one with only $Z$- detectors, and the complete Tanner graph that includes both detectors. The $4$- and $6$-cycles are two of the smallest cycles allowed on bipartite Tanner graphs.

From Table~\ref{tab:cycle-counts-bb} we see that the number of $4$-cycles and $6$-cycles drops for all biased noise models drops by an order of magnitude or more compared to the depolarizing model. Moreover, the number of short cycles in the $Z$-DEM, which is only affected by $X$-type errors, decrease from ancilla-only bias to non-CX bias to full-bias. Here $X$-type means any error that can flip a $Z$-detector and this trend is expected because the number of $X$ errors decreases from ancilla-only bias to non-CX bias to full-bias. 
Additionally, the number of short cycles in the $X$-DEM, affected by $Z$-type errors, remains the same for all the biased noise models. This is also expected because the number of $Z$-type error mechanisms, that is, error mechanisms that can flip a $X$-detector, remain the same across all the biased noise models.
Therefore, even when $Y$ errors on data qubits become improbable, the edges of the Tanner graph remain the same. 

Note that, a reduction in the number of cycles alone does not guarantee an improvement in the logical performance because the faults in the cycles do not happen with the same probability. 
It is possible that a code with fewer short cycles does worse than a code with more cycles if the probability of faults in these cycles is arranged less favorably for BP.
In fact, we will see in the next section that although $X$-DEMs have the same number of cycles across all three biased noise models, the logical performance of the full-bias case can be worse than that of non-CX biased or the ancilla-only biased noise because, for a fixed $p$, it has the highest probability of a $Z$-type error. 

Similarly to the gross code, we find that biasing the ancilla on the HGP code (Table~\ref{tab:cycle-counts-hgp}) reduces the number of short $4$-cycles compared to depolarizing noise.
The significant reduction in the number of these short cycles suggests that BP decoders may perform more accurately for biased noise models.
In the next section, we indeed see that this is the case; we present numerical results of circuit-level memory simulations on these codes for all four noise models.

\section{Results}
\label{sec:results}

In this section, we present results from circuit-level memory simulations of BB codes with distances $d=6,10,12$ under the four noise models.
As a case study for a code whose syndrome extraction circuit is guaranteed to be distance-preserving, we also study the performance of the $[[336, 20, 6]]$ HGP code.
Across these codes, we generally find that the ancilla-only biased noise model shows the best logical performance.

In our simulations, $d$ rounds of noisy syndrome extraction with noise models detailed in Section~\ref{sec:noise-models} are performed after noiseless state preparation, and the circuit is terminated by a round of noiseless stabilizer measurement.
The measurement outcomes of the $d$ noisy rounds and the final noiseless round are used to construct detector syndromes of the DEM.
This DEM and its syndromes are passed to a BP+OSD-CS-7 decoder~\cite{panteleev2021degenerate,roffe2020decoding, Roffe_LDPC_Python_tools_2022}, which uses a maximum number of $10,000$ BP iterations and OSD-CS processing with order $7$. 
We also compare two decoding strategies: one in which the $X$ and $Z$ syndromes are decoded separately and the other in which they are decoded together. 
We refer to these strategies as \textit{separate decoding} and \textit{joint decoding}, respectively.
We say a logical error occurs if the error followed by the inferred correction flip the value of any $X$ or $Z$ logical operators in the code block. 

\begin{figure}[htbp!]
    \centering
    \subfloat{
    \includegraphics[width=\linewidth]{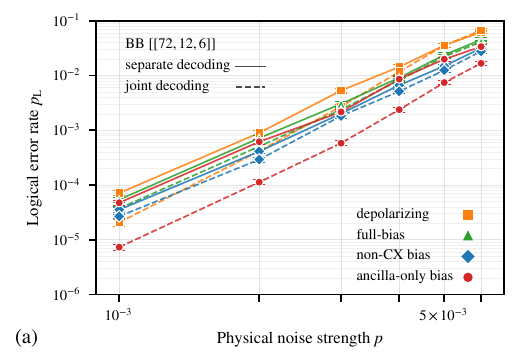}\label{fig:a}}\hfill
    \subfloat{
    \includegraphics[width=\linewidth]{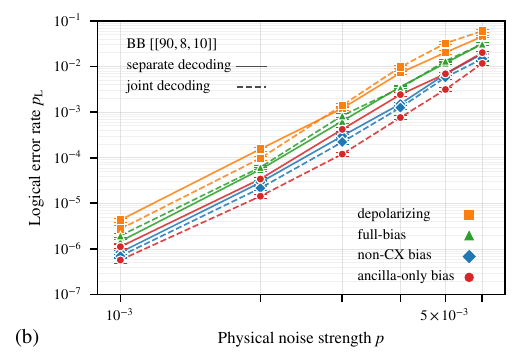}\label{fig:c}}\hfill
    \subfloat{
    \includegraphics[width=\linewidth]{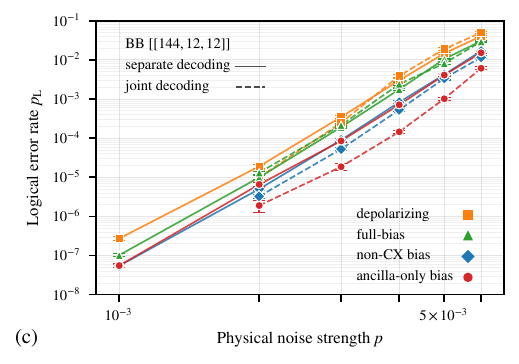}\label{fig:gross-code}}\hfill

  \caption{Logical error rates per round versus physical noise strength $p$ for all four noise models, depolarizing noise (orange), ancilla-only biased noise (red), non-CX biased noise (blue), and full-biased noise (green), on three BB codes: $[[72,12,6]]$ (a), $[[90,8,10]]$ (b), and $[[144,12,12]]$ (c).
  We also compare two decoding strategies: separate decoding (solid lines) and joint decoding (dashed lines).}
  \label{fig:bb-codes}
\end{figure}

\subsection{BB code results with infinite bias }Figure~\ref{fig:bb-codes} presents the logical error rate per syndrome extraction round for the $[[72,12,6]]$, $[[90,8,10]]$, and $[[144,12,12]]$ (gross) BB codes for different noise models at infinite bias. 
Across code distances, joint decoding (dashed lines) consistently matches or outperforms separate decoding (solid lines).
The gap in logical error rate between joint and separate decoding is especially large for the ancilla-only biased noise model. For example, for this noise at $p=2\times 10^{-3}$, the logical error rate under joint decoding is about a third of that at separate decoding.
The reason for this improved performance is that many correlated errors flip both $X$- and $Z$- detectors in the ancilla-only biased noise model, which the complete Tanner graph is able to capture. 
Here, the advantage of additional correlations in the combined Tanner graph is able to compensate for the increase in the number of short cycles (Table~\ref{tab:cycle-counts-bb}).
In contrast, for noise with fewer correlations across the two Tanner graphs, such as the full-biased noise model, joint decoding offers little to no advantage.

From Fig.~\ref{fig:bb-codes}, we observe that ancilla-only biased noise with joint decoding outperformed other noise models. As explained in the previous section, it does better than depolarizing noise because of the sharp decrease in the number of short cycles. 
On the other hand, the decrease in the short cycles from ancilla-only bias to non-CX bias to full-bias is not sufficient to compensate for the increase in the probability of these cycles. Consequently, the ancilla-only bias performs better than the other biased noise models. 

To probe the contribution of hook errors in the logical error rates, we evaluate 
the fraction of faults of weight $w$ that cause a failure and the error weight-distribution function for the gross code with separate decoding. After $10^{8}$ samples, we only find one logical failure for $w=12$, which indicates that a small weight error that also causes a logical failure is very rare.
Moreover, the weight distribution of errors at $p=2\times 10^{-3}$ is centered around $\sim 35$ for all noise models, and the probability to get a lower weight error is small in comparison. Consequently, at $p=2\times 10^{-3}$, we expect the logical error rate to be dominated by high-weight ($> d$) errors, and we conclude that the improvements in the logical performance around this error rate are not due to a reduction in hook errors. However, we expect low-weight hook errors to dominate at very low $p$.

\subsection{HGP code results with infinite bias} 
\begin{figure}[htbp!]
    \centering
    \includegraphics[width=1\linewidth]{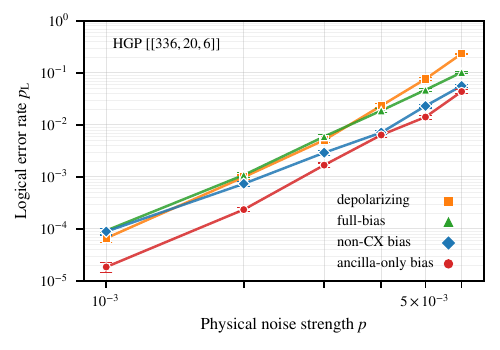}
    \caption{The $[[336, 20, 6]]$ HGP code for all four noise models under infinite bias ($\eta\rightarrow \infty$) and joint decoding, where $X$- and $Z$- syndromes are decoded together.}
    \label{fig:hgp-code}
\end{figure}

While the HGP codes have a distance-preserving syndrome extraction circuit, we nonetheless expect some improvement in the logical performance with biased noise due to a reduction in the short cycle count (Table~\ref{tab:cycle-counts-hgp}). 
In Fig.~\ref{fig:hgp-code}, we examine the performance of all four noise models under joint decoding for the $[[336, 20, 6]]$ cyclic HGP code.
We show performance only for joint decoding because its logical error rates were lower or matched the logical error rates for separate decoding.

As expected, we again find that the ancilla-only biased noise performs the best out of all the noise models under joint decoding.
An interesting observation is that the full-bias logical error rate (green) does not always outperform depolarizing noise (orange). This is because of two competing effects: the reduced short cycle count improves BP performance and lowers the logical error rate for biased noise, while the increased $Z$-type error probability causes BP to fail more on these cycles, increasing the logical error rate for full-biased noise. Which of the two effects wins depends on the code and $p$.

\begin{figure}[htbp!]
    \centering
    \includegraphics[width=1\linewidth]{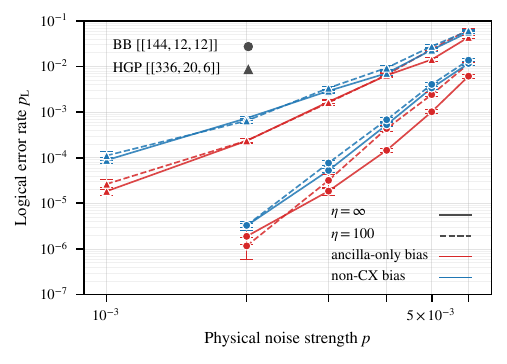}
    \caption{Comparing finite-bias ($\eta=100$, dashed lines) to infinite bias ($\eta\rightarrow \infty$, solid lines) simulations on the gross code (circles) and the $[[336,20,6]]$ HGP code (triangles). We consider ancilla-only biased noise (red) and non-CX biased noise (blue).}
    \label{fig:finite-bias}
\end{figure}
\subsection{Finite bias} For all simulations presented so far, we have used infinitely biased noise models with $\eta \rightarrow \infty$.
However, engineering qubits with infinite bias is unphysical. 
To establish that appreciable gains survive even at finite bias, we simulate the gross code and the $[[336,20,6]]$ cyclic HGP code at finite $\eta$ in Fig.~\ref{fig:finite-bias}.
We find that at a bias of $\eta=100$, the logical error rates are comparable to those of the infinitely biased case.
Moreover, for the gross code at a noise strength of $p=2\times 10^{-3}$, the logical error rate at $\eta=100$ is 8.5 times lower than for depolarizing noise.
This is promising evidence that the benefits of partially biased noise models persist in an experimentally realistic regime.

\section{Conclusion and Discussion}
\label{sec:conclusion}

In this work, we have shown that the logical performance of QLDPC codes can be improved if the ancilla noise is biased, without tailoring the code itself. By restricting the bias to the ancilla qubits, we suppress distance-reducing hook errors and reduce the number of short cycles which impair the performance of BP decoding. However, we find that the improvement in logical performance is mainly due to a reduction in short cycles for the noise strengths simulated in this paper in the range of $10^{-3}-10^{-2}$. 
For an experimentally realistic noise bias of $\eta=100$ at a physical error rate of $2\times 10^{-3}$, we find nearly an order of magnitude reduction in the logical error rate of the gross code with ancilla-only biased noise relative to unbiased noise. We observed similar overall trends using Relay-BP with separate decoding for the noise models considered here (but it performed slightly worse than joint decoding and therefore we didn't show the results)~\cite{muller2025improved}.
These improvements are consistent across the code families and decoders we studied, which suggests that biasing the ancilla may generally be a good strategy. These results are highly significant for practical QEC as they considerably ease biased noise requirements on the hardware and opens a new paradigm for biased noise QEC, which so far has been limited to bosonic qubits with difficult-to-engineer bias-preserving CX gates. 

There are several directions that follow naturally from this work. It would be interesting to see whether these noise model comparisons are flipped when using other decoders that are not as impacted by the presence of short cycles, such as beam-search decoders~\cite{ott2025decision,beni2025tesseract,leverrier2026approximating} or neural net decoders~\cite{blue2025machine,gu2026scalable}.
It would also be interesting to compare how bias-tailored codes with ancilla-only biased noise perform relative to non-CX biased and full-biased noise qubits. While we have only studied QLDPC memory performance here, how biased noise qubits can improve the performance of logical gates remains a largely open question that should be investigated, especially because it may be 
possible to overcome new decoding challenges that emerge during logical gates with biased noise.

\section*{Acknowledgments}

This work was supported by the U.S. Army Research Office (ARO) under grant W911NF-23-1-0051, and the U.S. Department of Energy, Office of Science, National Quantum Information Science Research Centers, and Codesign Center for Quantum Advantage (C2QA) under contract number DE-SC0012704.

\textit{Data availability---} the code used to generate the data for this work is provided at \url{https://github.com/rjb-14/biased-noise-ancilla-qldpc}.

\bibliography{references}

\end{document}